\documentclass[referee]{sn-jnl}

\usepackage{graphicx}%
\usepackage{multirow}%
\usepackage{amsmath,amssymb,amsfonts}%
\usepackage{amsthm}%
\usepackage{mathrsfs}%
\usepackage[title]{appendix}%
\usepackage{xcolor}%
\usepackage{textcomp}%
\usepackage{manyfoot}%
\usepackage{booktabs}%
\usepackage{algorithm}%
\usepackage{algorithmicx}%
\usepackage{algpseudocode}%
\usepackage{listings}%
\usepackage[numbers]{natbib}
\usepackage{hyperref}
\usepackage{tikz}
\usetikzlibrary{shapes.geometric, arrows.meta, positioning}
\begin{document}

	\title[Article Title]{Special Relativistic Kinematics from Wave Phase Coherence}

\author*[1]{\fnm{Emiliano} \sur{Puddu}}\email{epuddu@liuc.it}

\affil*[1]{\orgdiv{Department of Management Engineering}, \orgname{Università Cattaneo LIUC}, \orgaddress{\street{via Matteotti}, \city{Castellanza}, \postcode{21053}, \state{Va}, \country{Italy}}}
	
	\abstract{
		Special relativity is conventionally framed as a geometric theory of spacetime. Here, we shift the perspective, deriving its kinematical structure directly from the phase coherence of localized wave states. By assuming that physical propagation follows surfaces of constant phase and that matter possesses an internal rest-frame oscillation, we show that time dilation, energy, and momentum emerge naturally from the invariant accumulation of phase along a trajectory. In this view, proper time operationally counts the cycles of an internal clock, while the Minkowski interval becomes the necessary quadratic form to preserve phase invariance across inertial observers. Rather than introducing new dynamics, this approach bridges relativistic kinematics and wave propagation through a single, foundational principle.}
	
	\maketitle

	\section{Introduction}

Special relativity conventionally relies on the geometry of spacetime, encoding its foundational structure within the invariant Minkowski interval~\cite{Einstein1905,Minkowski1909}. Under this standard view, physical processes unfold across coordinate systems, and Lorentz transformations emerge strictly as the symmetries that preserve the metric tensor. Successful as this geometric formulation is, it does not represent the only logical starting point; the core objective of the present work is not to alter established formulas, but to invert the traditional logical hierarchy.
In several domains of physics---most notably wave mechanics and quantum theory---phase carries primary physical meaning, since observables are ultimately determined through interference and coherence phenomena~\cite{deBroglie1923,deBroglie1924,Schrodinger1926,Feynman1948}.
This operational reality suggests that relativistic kinematics can be entirely reframed by treating phase as the primary structural quantity.

For a relativistic wave defined by frequency $\omega$ and wavevector $\mathbf{k}$, the phase
\begin{equation}
	\Phi(\mathbf{x},t)=\mathbf{k}\cdot\mathbf{x}-\omega t
\end{equation}
can be written covariantly as $\Phi=p_\mu x^\mu/\hbar$, where $p_\mu$ is the energy-momentum four-vector under the $(+,-,-,-)$ metric signature. In this representation, the phase is a Lorentz scalar whose invariance reflects the consistency of wave propagation across all inertial frames. Although this property is textbook physics, its potential to serve as a foundational principle for kinematics is rarely made explicit.

In this work, we explore the consequences of adopting phase coherence as the core organizing principle of special relativity. We assume that physical propagation follows surfaces of constant phase and that localized states of matter possess an intrinsic oscillation characterized by a rest-frame frequency $\omega_0$~\cite{deBroglie1923,deBroglie1924,Dolce2013}. From this perspective, the accumulation of phase along a trajectory offers a natural operational definition of proper time. Specifically, the invariant phase increment reads
\begin{equation}
	d\Phi=-\omega_0\,d\tau,
\end{equation}
which identifies proper time $\tau$ simply as the parameter counting the internal cycles of a localized phase clock \cite{Muller2010}.

This single identification allows standard relativistic relations to emerge in a unified manner. Time dilation reflects the reduced rate of phase accumulation experienced by moving states, while energy and momentum follow directly from the de Broglie-Planck relations $E=\hbar\omega$ and $\mathbf{p}=\hbar\mathbf{k}$. Crucially, the Minkowski interval,
\begin{equation}
	ds^2=c^2dt^2-d\mathbf{x}^2,
\end{equation}
is no longer introduced as an independent geometric postulate; rather, it arises as the necessary quadratic structure required to preserve phase invariance across inertial observers. Spacetime geometry can therefore be interpreted as an encoding of the conditions that make relativistic phase coherence possible.

Within this same approach, the mass-frequency relation,
\begin{equation}
	\omega_0=\frac{mc^2}{\hbar},
\end{equation}
gains a direct physical meaning: mass represents the intrinsic oscillation scale of a localized state, while proper time tracks its phase history. Kinematics is thus completely recovered in wave-mechanical terms. While phase invariance is usually treated as a byproduct of covariance, here we treat it as the primary building block from which the kinematical structure of relativity naturally flows.

We emphasize that this formulation alters neither the predictions of special relativity nor its dynamical equations. Instead, it shifts the analytical focus toward phase invariance, providing a conceptual bridge toward more fundamental descriptions where spacetime geometry itself might emerge from underlying, non-geometric degrees of freedom.

The paper is structured to trace this logical shift. Section~II reviews the relativistic properties of phase and the requirements for frame-to-frame coherence. Section~III derives proper time from phase accumulation, while Section~IV establishes how the Minkowski interval arises from phase invariance. Section~V addresses the representation of energy, momentum, and mass. Finally, Section~VI and Section~VII evaluate the boundaries of this formulation and summarize our main conclusions.

\section{Phase invariance and relativistic coherence}

Let us focus on a localized wave state defined within a given inertial frame by the phase
\begin{equation}
	\Phi(\mathbf{x},t)=\mathbf{k}\cdot\mathbf{x}-\omega t.
\end{equation}
Here, surfaces of constant phase ($\Phi=\mathrm{const}$) map out the physical wavefronts. Operationally, these wavefronts mark the occurrence of propagation events.

For the wave description to remain consistent, different inertial observers must agree on the identity of these wavefronts. If $(\mathbf{x},t)$ and $(\mathbf{x}',t')$ represent the coordinates of the same physical event across two inertial frames, the underlying phase must map identically:
\begin{equation}
	\Phi(\mathbf{x},t)=\Phi'(\mathbf{x}',t').
\end{equation}
This identity defines the core requirement for relativistic phase coherence.

When inertial frames are linked by linear coordinate transformations, the phase must depend linearly on the spacetime coordinates to keep the wavefront mapping intact. Consequently, phase invariance demands that its coefficients transform contragrediently to the coordinates. This requirement frames the phase as a natural scalar contraction between a four-vector and the spacetime position:
\begin{equation}
	\Phi = k_\mu x^\mu,
\end{equation}
or equivalently,
\begin{equation}
	\Phi= -\frac{1}{\hbar}p_\mu x^\mu,
\end{equation}
where $k_\mu=p_\mu/\hbar$ denotes the wave four-vector and $p_\mu=(E/c,-\mathbf{p})$ represents the energy-momentum four-vector.

Expressed this way, the phase stands as a manifest Lorentz scalar, yielding the exact same value for any inertial observer tracking the event. Crucially, the physical substance lies not in the absolute phase value, but in phase differences, which ultimately dictate all interference and coherence phenomena. Phase invariance thereby provides the foundational structural condition for any consistent wave mechanics across moving frames.

\section{Proper time as phase accumulation}

Let us track the phase evolution along the trajectory of a localized wave packet. The differential of the phase reads
\begin{equation}
	d\Phi=\mathbf{k}\cdot d\mathbf{x}-\omega dt.
\end{equation}
This differential structure maps directly onto the Hamilton--Jacobi formulation of relativistic dynamics, where the phase acts as the classical action function \cite{LandauLifshitzFields}. Under the standard identification $\Phi = S/\hbar$, the phase increment along the path is given by $d\Phi = dS/\hbar = p_\mu dx^\mu/\hbar$.
Restricting our attention to the packet's actual worldline, we can set
\begin{equation}
	d\mathbf{x}=\mathbf{u}\,dt,
\end{equation}
where $\mathbf{u}$ denotes the group velocity. The phase increment along this trajectory then becomes
\begin{equation}
	d\Phi=(\mathbf{k}\cdot\mathbf{u}-\omega)\,dt.
\end{equation}

At this stage, we introduce our core physical postulate: any localized material state possesses an intrinsic rest-frame oscillation governed by a proper frequency $\omega_0$. This internal oscillation serves as a fundamental clock. We can define a scalar parameter $\tau$ by requiring that the phase accumulated along the trajectory be measured precisely in units of this rest-frame frequency:
\begin{equation}
	d\Phi=-\omega_0\,d\tau.
\end{equation}
Operationally, $\tau$ simply tracks the total cycle count of this internal phase clock.

Equating these two expressions for $d\Phi$ yields
\begin{equation}
	(\mathbf{k}\cdot\mathbf{u}-\omega)\,dt=-\omega_0\,d\tau.
\end{equation}
To evaluate the factor $(\omega-\mathbf{k}\cdot\mathbf{u})$, consider the state in its rest frame, where $\mathbf{k}_0=0$ and the phase evolves simply as
\begin{equation}
	\Phi_0(t_0)=-\omega_0 t_0.
\end{equation}
In an inertial frame where the packet moves with velocity $\mathbf{u}$, phase invariance dictates that the wave four-vector transforms via the standard Lorentz factor, giving
\begin{equation}
	\omega=\gamma\omega_0,
	\qquad
	\mathbf{k}=\gamma\frac{\omega_0}{c^2}\mathbf{u},
\end{equation}
where
\begin{equation}
	\gamma=\frac{1}{\sqrt{1-u^2/c^2}}.
\end{equation}
Direct substitution yields
\begin{align}
	\omega-\mathbf{k}\cdot\mathbf{u}
	&=
	\gamma\omega_0-\gamma\frac{\omega_0}{c^2}\mathbf{u}\cdot\mathbf{u}
	\nonumber\\
	&=
	\gamma\omega_0\left(1-\frac{u^2}{c^2}\right)
	\nonumber\\
	&=
	\frac{\omega_0}{\gamma}.
\end{align}
Inserting this result back into the phase increment equation, we obtain
\begin{equation}
	d\tau=\frac{dt}{\gamma}
	=
	dt\sqrt{1-\frac{u^2}{c^2}}.
\end{equation}

This result identifies $\tau$ precisely with the relativistic proper time. Rather than relying on a geometric postulate, proper time emerges here as the invariant parameter governing the phase accumulation of an internal clock. Time dilation thus gains a concrete, wave-mechanical interpretation: for a moving state, the internal phase accumulates more slowly relative to the laboratory time, meaning the internal clock runs slower.

\section{Invariant interval from phase coherence}

Let us now examine how phase invariance dictates the underlying geometry of spacetime. As established in the previous section, the phase increment along the trajectory of a localized state satisfies
\begin{equation}
	d\Phi = -\omega_0\, d\tau,
\end{equation}
where $\tau$ acts as the invariant parameter tracking the internal accumulation of phase.

Recalling the relation derived for $d\tau$, we have
\begin{equation}
	d\tau = dt \sqrt{1 - \frac{u^2}{c^2}},
\end{equation}
where $u = |\mathbf{u}|$ denotes the velocity of the localized state. Squaring this expression yields
\begin{equation}
	d\tau^2 = dt^2 \left(1 - \frac{u^2}{c^2}\right).
\end{equation}

By expressing the squared velocity as $u^2 = d\mathbf{x}^2/dt^2$, this relation straightforwardly rearranges into
\begin{equation}
	d\tau^2 = dt^2 - \frac{d\mathbf{x}^2}{c^2}.
\end{equation}
Multiplying through by $c^2$, we find
\begin{equation}
	c^2 d\tau^2 = c^2 dt^2 - d\mathbf{x}^2.
\end{equation}

This is the familiar Minkowski interval. Rather than serving as an independent geometric postulate, the interval emerges here as the exact quadratic structure required by, and compatible with, the invariant accumulation of phase.

Consequently, the spacetime interval can be reinterpreted as a physical encoding of the synchronization and coherence conditions across different inertial observers. Seen from this perspective, the Minkowski metric is not the foundational starting point of the theory, but rather the natural geometric representation of underlying phase invariance properties.

\section{Energy, momentum, and mass from phase structure}

Let us now examine the representation of energy, momentum, and mass within this phase-based approach. As established in Sec.~II, the phase admits a manifest covariant representation:
\begin{equation}
	\Phi=- \frac{1}{\hbar}p_\mu x^\mu,
\end{equation}
where $p_\mu$ denotes the energy-momentum four-vector.

Matching this expression against the explicit coordinate expansion,
\begin{equation}
	\Phi=\mathbf{k}\cdot\mathbf{x}-\omega t,
\end{equation}
straightforwardly yields the standard de Broglie--Planck relations:
\begin{equation}
	E=\hbar\omega,
	\qquad
	\mathbf{p}=\hbar\mathbf{k}.
\end{equation}
These identities map the kinematic parameters of the wave---frequency and wavevector---directly onto its dynamical properties.

Recall from Sec.~III that the phase increment along a localized worldline reads
\begin{equation}
	d\Phi=-\omega_0\,d\tau,
\end{equation}
where $\omega_0$ sets the invariant, rest-frame frequency scale of the state. By defining the corresponding rest energy as
\begin{equation}
	E_0=\hbar\omega_0,
\end{equation}
the fundamental mass-frequency relation follows naturally:
\begin{equation}
	m=\frac{E_0}{c^2}=\frac{\hbar\omega_0}{c^2}.
\end{equation}

Furthermore, the transformation properties of the wave four-vector ensure that for a state moving at velocity $\mathbf{u}$, the energy and momentum take their familiar relativistic forms:
\begin{equation}
	E=\gamma mc^2,
	\qquad
	\mathbf{p}=\gamma m\mathbf{u},
\end{equation}
where
\begin{equation}
	\gamma=\frac{1}{\sqrt{1-u^2/c^2}}.
\end{equation}
As expected, these dynamical quantities obey the invariant dispersion relation:
\begin{equation}
	E^2=p^2c^2+m^2c^4.
\end{equation}

From this perspective, mass gains a transparent physical meaning as the intrinsic oscillation scale of the state in its rest frame. While proper time tracks the internal accumulation of these phase cycles, energy and momentum characterize the frequency and spatial modulation of the wave as viewed from an arbitrary inertial frame. Relativistic kinematics is thus fully recovered as a coherent, unified manifestation of the underlying phase structure.

	\section{Discussion}

The phase-based reconstruction developed in this work offers an alternative derivation of relativistic kinematics, demonstrating how proper time, energy, momentum, and the invariant interval naturally flow from the phase coherence of localized wave states. Because the resulting structures map identically onto standard special relativity, this approach leaves all empirical predictions unaltered.

Crucially, this formulation does not propose new dynamics. Its value lies instead in a structural reinterpretation that elevates phase invariance to a primary organizing principle. Within this logical inversion, the Minkowski interval ceases to be an independent geometric axiom; it emerges as the exact quadratic form required to sustain the invariant accumulation of phase along physical worldlines.

Interpreting proper time as the cycle count of an intrinsic oscillation builds a natural bridge to the wave-mechanical description of matter. Specifically, linking mass directly to a rest-frame frequency aligns seamlessly with the historical insights of de Broglie and Compton, highlighting how internal periodicity underpins relativistic behavior. Furthermore, this perspective anchors kinematic properties directly to interference phenomena~\cite{Feynman1948}, where observable effects are fundamentally driven by phase differences. The present approach can thus be viewed as an operational link between the geometric scaffolding of spacetime and the phase-centric formalism of quantum theory.

Naturally, certain boundaries define the scope of this work. Our analysis remains strictly kinematical; we do not address the dynamical origin of the intrinsic restfrequency, nor do we solve for the underlying field equations that govern wave propagation. Additionally, the framework is currently confined to inertial observers, leaving extensions to accelerated frames or gravitational fields open for future research.

Nevertheless, shifting the foundational emphasis to phase provides a self-consistent, unified reading of relativistic kinematics. By demonstrating that spacetime geometry can be viewed as an encoding of phase coherence, this perspective may offer a fresh vantage point for exploring the intersection of wave propagation, metric structures, and emergent theories of spacetime. This includes potential extensions toward effective optical metrics or weak-field gravitational analogs, though such avenues remain beyond our immediate scope.
	
	\section{Conclusion}

This paper explores an alternative formulation of special relativity, showing that its kinematical architecture can be systematically derived from the phase coherence of localized wave states. By elevating phase to the primary invariant quantity of the theory, we have demonstrated that proper time operationally counts the cycles of an intrinsic oscillation, while the Minkowski interval emerges naturally as the precise quadratic form required to preserve this phase invariance across frames.

Consequently, dynamic properties such as energy and momentum flow directly from the underlying wave characteristics, allowing mass to be reinterpreted as the fundamental rest-frame frequency of an internal clock. Rather than introducing kinematics through a series of rigid geometric postulates, these results offer a unified, self-consistent perspective where spacetime intervals are understood as an operational manifestation of phase evolution.

While this formulation leaves the empirical predictions of special relativity entirely unchanged, it clarifies its structural foundations by highlighting phase coherence as a powerful organizing principle. This approach may provide fresh insights wherever wave-mechanical and geometric descriptions intersect, offering a promising baseline for future work aimed at extending these principles to more generalized physical settings.

\begin{figure}[htbp]
\centering
\begin{tikzpicture}[
    node distance=0.9cm and 1.2cm,
    block/.style={rectangle, draw, fill=blue!5, text width=4.0cm, align=center, rounded corners, minimum height=1.1cm, font=\small},
    myblock/.style={rectangle, draw, fill=orange!5, text width=4.0cm, align=center, rounded corners, minimum height=1.1cm, font=\small},
    arrow/.style={-Stealth, thick}
]

    \node (std_title) [font=\bfseries] {Standard Geometric Approach};

    \node (mink) [block, below=0.4cm of std_title] {Minkowski Spacetime Postulate ($ds^2$)~\cite{Einstein1905,Minkowski1909}};
    \node (lorentz) [block, below=of mink] {Lorentz Transformations};
    \node (cov) [block, below=of lorentz] {Covariance of Wave Equations};
    \node (phase_inv) [block, below=of cov] {Phase Invariance ($\Phi$) as a Consequence};

    \draw [arrow] (mink) -- (lorentz);
    \draw [arrow] (lorentz) -- (cov);
    \draw [arrow] (cov) -- (phase_inv);


    \node (your_title) [font=\bfseries, right=2.0cm of std_title] {Present Phase-Based Approach};

    \node (phase_coherence) [myblock, below=0.4cm of your_title] {Phase Coherence of Localized States ($\Phi$)};
    \node (proper_time) [myblock, below=of phase_coherence] {Proper Time as Phase Accumulation ($\tau$)~\cite{LandauLifshitzFields}};
    \node (mass_freq) [myblock, below=of proper_time] {Emergence of Minkowski Metric ($ds^2$)};
    \node (mink_emerge) [myblock, below=of mass_freq] {Mass from Intrinsic Rest Frequency ($\omega_0$)};

    \draw [arrow] (phase_coherence) -- (proper_time);
    \draw [arrow] (proper_time) -- (mass_freq);
    \draw [arrow] (mass_freq) -- (mink_emerge);


    \path (lorentz) -- (proper_time) node[midway, font=\itshape\bfseries\small, align=center, fill=white, inner sep=4pt, yshift=-0.4cm] {Logical\\Inversion};

\end{tikzpicture}
\caption{Comparison of logical hierarchies in special relativity. Left: The traditional geometric view, which treats phase invariance as a structural consequence of the underlying spacetime metric. Right: The present wave-mechanical approach, where the quadratic Minkowski interval emerges from foundational phase coherence conditions.}
\label{fig:logical_hierarchy}
\end{figure}
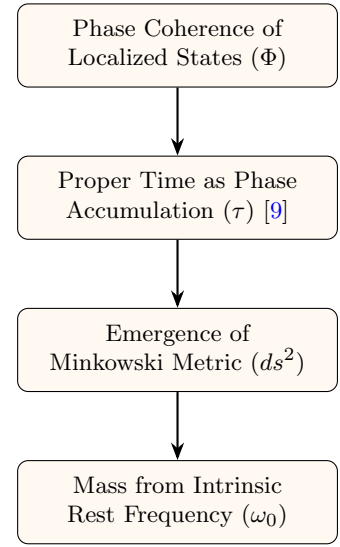

	\bibliographystyle{sn-mathphys-num}
	\bibliography{references}

@article{Einstein1905,
	author = {Einstein, Albert},
	title = {Zur Elektrodynamik bewegter K{\"o}rper},
	journal = {Annalen der Physik},
	volume = {17},
	pages = {891--921},
	year = {1905},
	doi =  {10.1002/andp.19053221004}
}

@article{Minkowski1909,
    author  = {Minkowski, Hermann},
    title   = {Raum und Zeit},
    journal = {Jahresbericht der Deutschen Mathematiker-Vereinigung},
    volume  = {18},
    pages   = {75--88},
    year    = {1909},
    url     = {http://eudml.org/doc/145167}
}

@article{deBroglie1923,
	author = {de Broglie, Louis},
	title = {Ondes et quanta},
	journal = {Comptes Rendus de l'Acad{\'e}mie des Sciences},
	volume = {177},
	pages = {507--510},
	year = {1923}
}

@phdthesis{deBroglie1924,
	author = {de Broglie, Louis},
	title = {Recherches sur la th{\'e}orie des quanta},
	school = {Universit{\'e} de Paris, Paris},
	year = {1924}
}

@article{Feynman1948,
    author  = {Feynman, Richard P.},
    title   = {Space-Time Approach to Non-Relativistic Quantum Mechanics},
    journal = {Reviews of Modern Physics},
    volume  = {20},
    number  = {2},
    pages   = {367--387},
    year    = {1948},
    doi     = {10.1103/RevModPhys.20.367}
}

@book{LandauLifshitzFields,
	author	 = {Landau, Lev D. and Lifshitz, Evgeny M.},
	title	 = {The Classical Theory of Fields},
	edition	 = {4th},
	publisher	 = {Pergamon Press},
	address = {Oxford},
	year = {1975},
	doi = {10.1016/c2009-0-14608-1}
}

@article{Schrodinger1926,
    author  = {Schr{\"o}dinger, Erwin},
    title   = {An Undulatory Theory of the Mechanics of Atoms and Molecules},
    journal = {Physical Review},
    volume  = {28},
    number  = {6},
    pages   = {1049--1070},
    year    = {1926},
    doi     = {10.1103/PhysRev.28.1049}
}

@article{Dolce2013,
    author  = {Dolce, Donatello},
    title   = {Intrinsic Periodicity: The Forgotten Lesson of Quantum Mechanics},
    journal = {Journal of Physics: Conference Series},
    volume  = {442},
    number  = {1},
    pages   = {012048},
    year    = {2013},
    doi     = {10.1088/1742-6596/442/1/012048}
}

@article{Muller2010,
    author = {M{\"u}ller, Holger and Peters, Achim and Chu, Steven},
    title = {A Precision Measurement of the Gravitational Redshift by the Interference of Matter Waves},
    journal = {Nature},
    volume = {463},
    pages = {926--929},
    year = {2010},
    doi = {10.1038/nature08776}
}

\end{document}